\documentclass[cits]{PoS}
\usepackage{psfig,epsfig,colordvi}
\def\slashed{{/}\mskip-10.0mu}

\title{Two-loop renormalization of fermion bilinear operators on the lattice}

\ShortTitle{Two-loop renormalization of fermion bilinear operators on the lattice}

\author{A.~Skouroupathis\thanks{Work supported in part by
    the Research Promotion Foundation of Cyprus (Proposal Nr: $\rm
    ENI\Sigma X$/0506/17).}~~and \speaker{H.~Panagopoulos} \\
        Department of Physics, University of Cyprus  \\
        E-mail: \email{php4as01@ucy.ac.cy}, \email{haris@ucy.ac.cy}}

\abstract{We compute the renormalization functions on the
  lattice, in the $RI^{\,\prime}$ scheme, of local bilinear quark
  operators $\bar{\psi}\Gamma\psi$, where $\Gamma= \hat{1},\,
  \gamma_5,\,\gamma_{\mu},\,\gamma_5\,\gamma_{\mu},\,\gamma_5\,
  \sigma_{\mu\,\nu}$. This calculation is carried out to two loops for
  the first time. We consider both the flavor non-singlet and singlet
  operators.    

As a prerequisite for the above, we compute the quark field 
renormalization, $Z_{\psi}^{L,RI^{\,\prime}}$, up to two loops. We also compute the
1-loop renormalization functions for the gluon field,
$Z_A^{L,RI^{\,\prime}}$, ghost field, $Z_c^{L,RI^{\,\prime}}$, gauge
parameter, $Z_\alpha^{L,RI^{\,\prime}}$, and coupling constant
$Z_g^{L,RI^{\,\prime}}$.   

We use the clover action for fermions and the Wilson action for
gluons. Our results are given as an explicit function of the coupling
constant $a_\circ =g_\circ^2/16\pi^2$, the clover coefficient
$c_{SW}$, and the number of fermion colors ($N_c$) and flavors
($N_f$), in the renormalized Feynman gauge. All 1-loop quantities are
evaluated in an arbitrary gauge. 

Finally, we present our results in the $\overline{MS}$ scheme, for
easier comparison with calculations in the continuum. 
We have generalized to fermionic fields in an arbitrary
representation. Some special features of superficially divergent
integrals, obtained from the evaluation of two-loop Feynman diagrams, 
are presented in detail in Ref. \cite{SP1}.}

\FullConference{The XXVII International Symposium on Lattice Field Theory\\
                 July 26-31, 2009\\
                 Peking University, Beijing, China}

\begin{document}

\section{Introduction}

Numerical simulations of QCD, formulated on the lattice, make use of a 
variety of composite operators, made out of quark fields. Matrix
elements and correlation functions of a whole variety of such
operators, are computed in order to study hadronic properties in this
context. A proper renormalization of these operators is essential for
the extraction of physical results from the lattice. 

In this work we study the renormalization function, $Z_\Gamma$, of
fermion bilinears  ${\cal O}=\bar{\psi}\Gamma\psi$ on the lattice,
where $\Gamma =\hat{1},\,\gamma_5,\,\gamma_{\mu},\,\gamma_5\,
\gamma_{\mu},\,\gamma_5\,\sigma_{\mu\,\nu}$
($\sigma_{\mu\,\nu}=1/2\,[\gamma_\mu,\gamma_\nu]$). We consider both
flavor singlet and nonsinglet operators. We employ the standard Wilson
action for gluons and clover-improved Wilson fermions. The number of
quark flavors $N_f$, the number of colors $N_c$ and the clover
coefficient $c_{{\rm SW}}$ are kept as free parameters. One necessary
ingredient for the renormalization of fermion bilinears is the 2-loop
quark field renormalization, $Z_{\psi}$, calculated in \cite{SP2}. The
one-loop expression for the renormalization function 
$Z_g$ of the coupling constant is also necessary for expressing the 
results in terms of both the bare and the renormalized coupling
constant.

Our two-loop calculations have been performed in the bare and in 
the renormalized Feynman gauge. For the latter, we need the 1-loop 
renormalization functions $Z_\alpha$ and $Z_A$ of the gauge
parameter and gluon field respectively, as well as the one-loop
expressions for $Z_\Gamma$ with an arbitrary value of the gauge
parameter.

The main results presented in this work are 2-loop bare Green's
functions (amputated, one-particle irreducible (1PI)), for the scalar,
pseudoscalar, vector, axial vector and tensor operator, as functions
of the lattice spacing, $a_{_{\rm L}}$, and the external momentum
$q$. In general, one can use bare Green's functions to construct 
$Z_{{\cal O}}^{X,Y}$, the renormalization function for operator ${\cal
  O}$, computed within a regularization $X$ and renormalized in a
scheme $Y$. We employ two widely used schemes to compute the various 
2-loop renormalization functions: The $RI^{\prime}$ scheme and the
$\overline{MS}$ scheme. 

The present work is the first two-loop computation of the
renormalization of fermion bilinears on the lattice. One-loop
computations of the same quantities exist for quite some time now
(see, e.g., \cite{Capitani} and references
therein). There have been made several attempts to estimate $Z_{{\cal
    O}}$ non-perturbatively; recent results can be found in
Refs. \cite{Aoki,Galletly,DellaMorte}. A series
of results have also been obtained using stochastic perturbation
theory \cite{DiRenzo}. A related computation,
regarding the fermion mass renormalization $Z_m$ with staggered
fermions, can be found in \cite{Trottier}.

\section{Formulation of the problem}

We will make use of the Wilson formulation of the QCD action on the
lattice, with the addition of the clover (SW)
term for fermions. In standard notation, it reads:
\begin{eqnarray}
S_L &=& S_G + \sum_{f}\sum_{x} (4r+m_{\rm o})\bar{\psi}_{f}(x)\psi_f(x)
\nonumber \\
&-& {1\over 2}\sum_{f}\sum_{x,\,\mu}
\bigg{[}\bar{\psi}_{f}(x) \left( r - \gamma_\mu\right)
U_{x,\,x+\mu}\,\psi_f(x+{\mu}) 
+\bar{\psi}_f(x+{\mu})\left( r + \gamma_\mu\right)
U_{x+\mu,\,x}\,\psi_{f}(x)\bigg{]} \\
&-& {1\over 4}\,c_{\rm SW}\,\sum_{f}\sum_{x,\,\mu,\,\nu} \bar{\psi}_{f}(x)
\,\sigma_{\mu\nu} \,{\hat F}_{\mu\nu}(x) \,\psi_f(x), \nonumber
\label{latact} 
\end{eqnarray}
\begin{eqnarray}
{\rm where:}\ {\hat F}_{\mu\nu} &\equiv& {1\over{8a^2}}\,
(Q_{\mu\nu} - Q_{\nu\mu})\\
{\rm and:} \ Q_{\mu\nu} &=& U_{x,\, x+\mu}\,U_{x+\mu,\, x+\mu+\nu}\,U_{x+\mu+\nu,\, x+\nu}
\,U_{x+\nu,\, x}+ U_{ x,\, x+ \nu}\,U_{ x+ \nu,\, x+ \nu- \mu}\,U_{ x+ \nu- \mu,\, x- \mu}\,U_{ x- \mu,\, x} \nonumber \\
&+& U_{ x,\, x- \mu}\,U_{ x- \mu,\, x- \mu- \nu}\,U_{ x- \mu- \nu,\, x- \nu}\,U_{ x- \nu,\, x}+ U_{ x,\, x- \nu}\,U_{ x- \nu,\, x- \nu+ \mu}\,U_{ x- \nu+ \mu,\, x+ \mu}\,U_{ x+ \mu,\, x}\qquad\,
\label{latact2}
\end{eqnarray}

$S_G$ is the standard pure gluon action, made out of $1{\times}1$
plaquettes. $r$ is the Wilson parameter (set to $r=1$ henceforth); 
$f$ is a flavor index. Powers of the lattice spacing $a_{_{\rm L}}$
have been omitted and may be directly reinserted by dimensional
counting.   

The ``Lagrangian mass'' $m_{\rm o}$ is a free parameter in principle. However,
since we will be using mass independent renormalization schemes, all
renormalization functions which we will be calculating, must be
evaluated at vanishing renormalized mass, that is, when $m_{\rm o}$ is
set equal to the critical value $m_{\rm cr}$: $m_{\rm o}\to m_{\rm
  cr}=m_1\,g_\circ^2+{\cal O}(g_\circ^4)$.

One prerequisite to our programme consists of the renormalization
functions, $Z_A$, $Z_c$, $Z_\psi$, $Z_g$ and $Z_\alpha$, for the
gluon, ghost and fermion fields ($A_\mu^a,\ c^a,\ \psi$), and for the
coupling constant $g$ and gauge parameter $\alpha$, respectively (for
definitions of these quantities, see Ref. \cite{SP2}); we will also
need the fermion mass counterterm $m_{\rm cr}$. These quantities are
all needed to one loop, except for $Z_\psi$ which is required to two
loops.  The value of each $Z_{{\cal O}}$ depends both on the
regularization $X$ and on the renormalization scheme $Y$ employed, and
thus should properly be denoted as $Z^{X,Y}_{{\cal O}}$.

As mentioned before, we employ the $RI^\prime$ renormalization scheme
\cite{Martinelli}, which is more immediate for a
lattice regularized theory. It is defined by imposing a set of
normalization conditions on matrix elements at a scale $\bar{\mu}$,
where (just as in the $\overline{MS}$ scheme): 
\begin{equation}
\bar{\mu}=\mu\,(4\pi/e^{\gamma_{_{\,\rm E}}})^{1/2} 
\label{mubar}
\end{equation}
where $\gamma_{_{\,\rm E}}$ is the Euler constant and $\mu$ is the scale
entering the bare coupling constant $g_\circ = \mu^\epsilon\,Z_g\,g$
when regularizing in $D=4-2\epsilon$ dimensions.

\section{Renormalization of fermion bilinears}

The lattice operators ${\cal O}_{\Gamma}=\bar{\psi}\,\Gamma\,\psi$
must, in general, be renormalized in order to have finite matrix
elements. We define renormalized operators by 
\begin{equation}
{\cal
  O}^{RI^{\prime}}_{\Gamma}=Z^{L,RI^{\prime}}_{\Gamma}(a_{_{\rm L}}\bar\mu)\,{\cal
  O}_{\Gamma\,{\rm o}}
\label{RenormOper}
\end{equation}

The renormalization functions $Z_{\Gamma}^{L,RI^{\prime}}$ can be
extracted through the corresponding bare 2-point functions
$\Sigma^L_{\Gamma}(q a_{_{\rm L}})$ (amputated, 1PI) on the lattice,
through the employment of the $RI^{\prime}$ renormalization
conditions: 
\begin{equation}
\lim_{a_{_{\rm L}}\rightarrow 0}\left[Z_{\psi}^{L,RI^{\,\prime}}\,Z_\Gamma^{L,RI^{\,\prime}}\,
\Sigma^L_\Gamma(q a_{_{\rm L}})\right]_{q^2=\bar{\mu}^2} = \Gamma_{\rm
tree},
\label{ZGammaRule}
\end{equation}
where $\Sigma^L_\Gamma(q a_{_{\rm L}})$ is the appropriate
bare 1PI 2-point Green's function on the lattice and $\Gamma_{\rm
  tree}$ is its tree-level value.
For the vector (V), axial-vector (AV) and tensor (T) operators, we can
express the bare Green's functions in the following way: 
\begin{eqnarray}
\Sigma^L_V(q a_{_{\rm L}})&=&\gamma_\mu\,\Sigma^{(1)}_V (q a_{_{\rm L}}) +
\frac{q^\mu\slashed q}{q^2}\Sigma^{(2)}_V (q a_{_{\rm L}}) \nonumber \\ 
\Sigma^L_{AV}(q a_{_{\rm L}})&=&\gamma_5\gamma_\mu\,\Sigma^{(1)}_{AV} (q a_{_{\rm L}}) + 
\gamma_5\frac{q^\mu\slashed q}{q^2}\Sigma^{(2)}_{AV} (q a_{_{\rm L}}) \label{2ptFunct}
\\
\Sigma^L_T(q a_{_{\rm L}})&=&\gamma_5\,\sigma_{\mu\,\nu}\Sigma^{(1)}_T(q a_{_{\rm L}}) + 
\gamma_5\frac{\slashed q\,(\gamma_\mu q_\nu - \gamma_\nu
  q_\mu)}{q^2}\Sigma^{(2)}_T(q a_{_{\rm L}})  \nonumber 
\label{brgrnfunctn}
\end{eqnarray}
Only the $\Sigma^{(1)}_{V,\,AV,\,T}(q a_{_{\rm L}})$ parts are involved
in Eq. (\ref{ZGammaRule}). It is worth noting here that terms which
break Lorentz invariance (but are compatible with hypercubic
invariance), such as $\gamma_\mu\,(q^\mu)^2/q^2$, turn out to be
absent from all bare Green's functions; thus, the latter have the same
Lorentz structure as in the continuum.

For easier comparison with calculations coming from the continuum, we
need to express our results in the $\overline{MS}$ scheme. For each
renormalization function on the lattice, $Z^{L,RI^{\prime}}_{{\cal
    O}}$, we can construct its $\overline{MS}$ counterpart using  
conversion factors:
\begin{equation}
C_\Gamma(g,\alpha)\equiv\frac{Z_\Gamma^{L,RI^{\prime}}}{Z_\Gamma^{L,\overline{MS}}}=
\frac{Z_\Gamma^{DR,RI^{\prime}}}{Z_\Gamma^{DR,\overline{MS}}}
\label{CGamma}
\end{equation}
These conversion factors are regularization independent; thus they can
be calculated more easily in dimensional (DR), rather than Lattice (L),
regularization, (see, e.g., Ref. \cite{Gracey}). Due to the
non-unique generalization of $\gamma_5$ to D dimensions, the
pseudoscalar and axial-vector bilinear operators require special
attention in the $\overline{MS}$ scheme. 

For a more detailed analysis of the renormalization of fermion
bilinears and their conversion to the $\overline{MS}$ scheme, see
Refs. \cite{SP1,SP2}.  

\section{Computation and Results}

\begin{figure}[h]
\centerline{\includegraphics[scale=0.36]{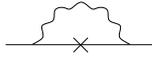}}
\caption{One-loop diagram contributing to $Z_\Gamma$. A
  wavy (solid) line represents gluons (fermions). A cross denotes the
  Dirac matrices $\hat{1}$ (scalar), $\gamma_{5}$ (pseudoscalar),
  $\gamma_\mu$ (vector), $\gamma_5\gamma_\mu$ (axial vector) and
  $\gamma_{5}\sigma_{\mu\nu}$ 
  (tensor). \label{ZVAT1loopDiagrams}}  
\end{figure}

\begin{figure}[b]
\centerline{\psfig{figure=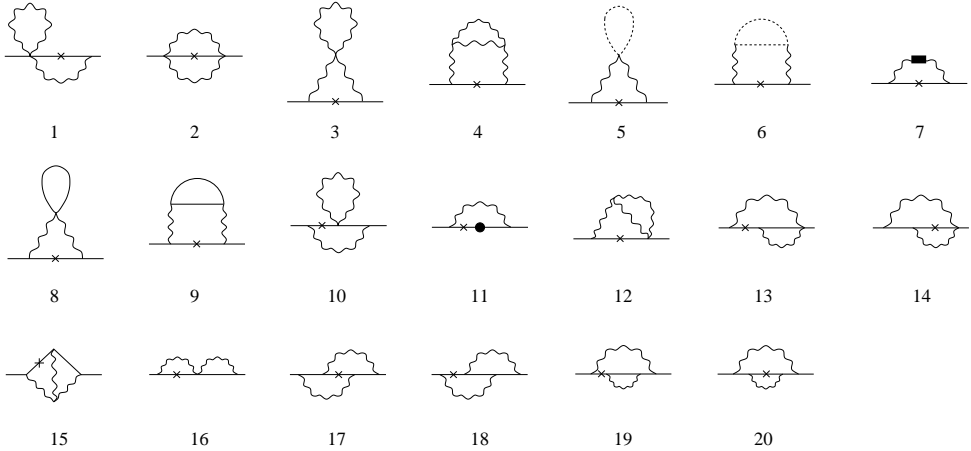,scale=0.25}}
\caption{Two-loop diagrams contributing to $Z_\Gamma$. 
  Wavy (solid, dotted) lines represent gluons (fermions, ghosts). A
  solid box denotes a vertex from the measure part of the action; a
  solid circle is a mass counterterm; crosses denote the Dirac matrices
  $\hat{1}$ (scalar), $\gamma_{5}$ (pseudoscalar), $\gamma_\mu$
  (vector), $\gamma_5\gamma_\mu$ (axial-vector) and
  $\gamma_{5}\sigma_{\mu\nu}$ (tensor). \label{ZVAT2loopDiagrams}}   
\end{figure}

\begin{figure}
\centerline{\psfig{figure=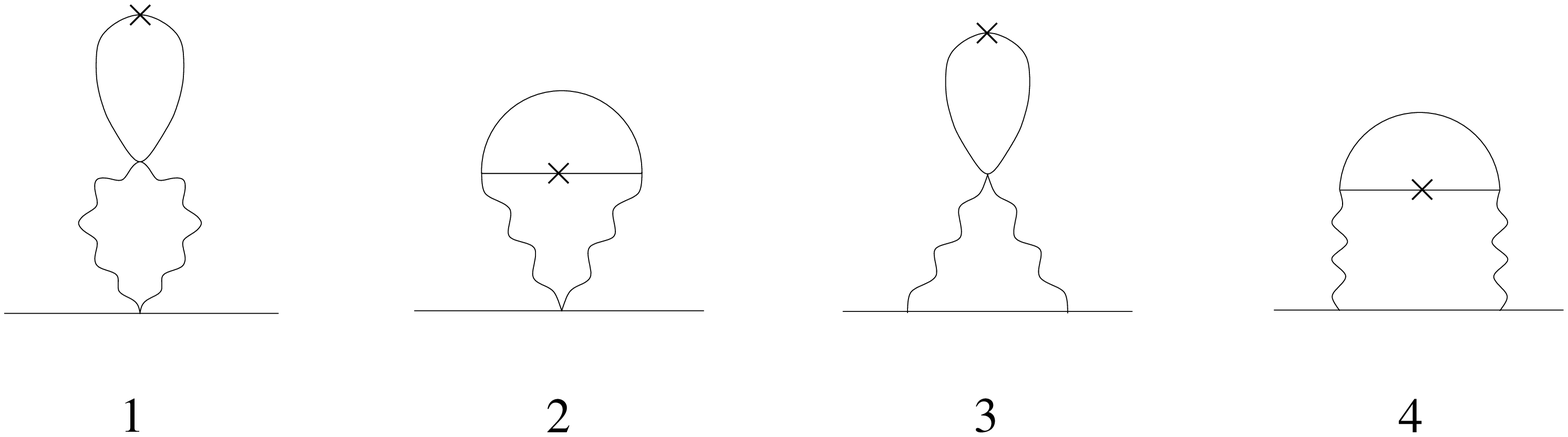,height=2truecm}}
\caption{Extra two-loop diagrams contributing to $Z_{S,\,singlet}$ and
  $Z_{AV,\,singlet}$. A cross denotes an insertion of a flavor singlet
  operator. Wavy (solid) lines represent gluons
  (fermions). \label{singletVAT}}   
\end{figure}

The Feynman diagrams contributing to the bare Green's functions, at 1-
and 2-loop level, are shown in Figs. \ref{ZVAT1loopDiagrams} and
\ref{ZVAT2loopDiagrams}, respectively. For flavor singlet bilinears,
there are 4 extra diagrams, shown in Fig. \ref{singletVAT}, which
contain the operator insertion inside a closed fermion loop. These
diagrams give a nonzero contribution only in the scalar and
axial-vector cases. 

In Figs. \ref{ZVAT1loopDiagrams} to \ref{singletVAT}, ``mirror''
diagrams (those in which the direction of the external fermion line is
reversed) should also be included. In most cases, these coincide
trivially with the original diagrams; even in the remaining cases,
they can be seen to give equal contribution, by invariance under
charge conjugation.

The evaluation of all Feynman diagrams leads directly to
the corresponding bare Green's functions $\Sigma^L_\Gamma$. These, in
turn, can be converted to the corresponding renormalization functions
$Z_\Gamma^{L,RI^{\,\prime}}$, via Eq. (\ref{ZGammaRule}). One-loop 
results for $Z_\Gamma^{L,RI^{\prime}}$ are presented below in a 
generic gauge. The errors appearing in these expressions, result from
an extrapolation to infinite lattice.        
\begin{eqnarray}
Z_S^{L,RI^{\prime}}= 1 + \frac{g_\circ^2}{16\pi^2}\,c_F &\Big{[}&3\,\ln(a_{_{\rm L}}^2 \bar{\mu}^2)- \alpha_\circ - 16.9524103(1) \nonumber \\
&& \,\, - 7.7379159(3)\,c_{{\rm SW}} + 1.38038065(4)\,c_{{\rm SW}}^2 \Big{]} \label{ZS1loopRI} \\
Z_P^{L,RI^{\prime}}= 1 + \frac{g_\circ^2}{16\pi^2}\,c_F&\Big{[}&3\,\ln(a_{_{\rm L}}^2 \bar{\mu}^2)- \alpha_\circ - 26.5954414(1) \nonumber \\
&&\,\, + 2.248868528(3)\,c_{{\rm SW}} - 2.03601561(4)\,c_{{\rm SW}}^2
  \Big{]} \label{ZP1loopRI} \\
Z_T^{L,RI^{\prime}}= 1 + \frac{g_\circ^2}{16\pi^2}\,c_F &\Big{[}&-\ln(a_{_{\rm L}}^2 \bar{\mu}^2)+\alpha_\circ -17.018079209(7) \nonumber \\
&& \,\, +3.91333261(4)\,c_{{\rm SW}} + 1.972295300(5)\,c_{{\rm SW}}^2 \Big{]} \label{ZT1loopRI}
\end{eqnarray}
The corresponding expressions for $Z_V^{L,RI^\prime}$,
$Z_{AV}^{L,RI^\prime}$ can be read off from
Eqs. (\ref{ZV2loopRI},\ref{ZA2loopRI}) below. 

\begin{figure}[h]
\centerline{\psfig{figure=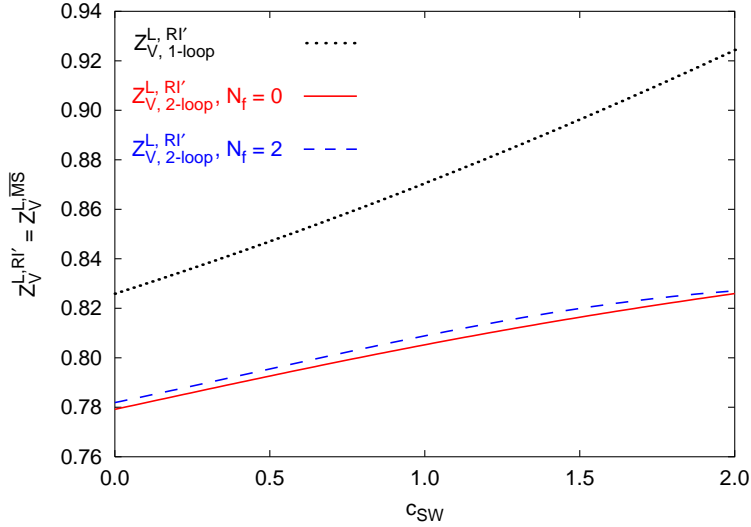,scale=0.4, angle=-90}}
\caption{$Z_V^{L,RI^{\,\prime}}(a_{_{\rm L}}\bar{\mu}) =
  Z_V^{L,\overline{MS}}(a_{_{\rm L}}\bar{\mu})$ versus $c_{\rm SW}$
  ($N_c=3$, $\bar{\mu}=1/a_{_{\rm L}}$, $\beta_\circ = 6.0$). Results
  up to 2 loops are shown for $N_f=0$ (solid line) and $N_f=2$
  (dashed line); one-loop results are plotted with a dotted
  line. \label{ZVRIplot}}    
\end{figure}

\begin{figure}[h]
\centerline{\psfig{figure=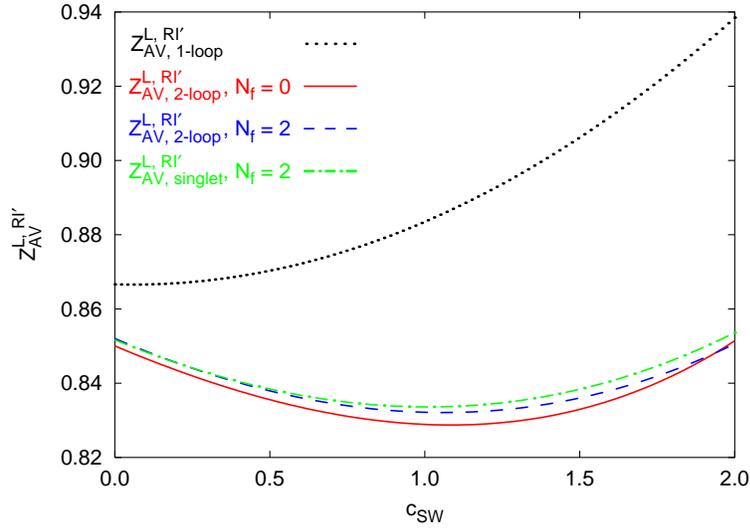,scale=0.4,angle=-90}}
\caption{$Z_{AV}^{L,RI^{\,\prime}}(a_{_{\rm L}}\bar{\mu})$ versus $c_{\rm SW}$
  ($N_c=3$, $\bar{\mu}=1/a_{_{\rm L}}$, $\beta_\circ = 6.0$). Results
  up to 2 loops, for the flavor nonsinglet operator, are shown for
  $N_f=0,\,2$ (solid line, dashed line); 2-loop results for
  the flavor singlet operator, for $N_f=2$, are plotted with a
  dash-dotted line; one-loop results are plotted with a dotted
  line. \label{ZARIplot}}    
\end{figure}

We present below $Z_V^{L,RI^{\prime}}$ and $Z_{AV}^{L,RI^{\prime}}$ to
two loops in the renormalized Feynman gauge. The corresponding plots
are exhibited in Figs. \ref{ZVRIplot} and \ref{ZARIplot}, as functions
of the clover parameter, $c_{\rm SW}$. 
For a complete set of results, regarding the
renormalization functions and renormalized Green's functions, both in
the $RI^\prime$ and in the $\overline{MS}$ scheme, the reader should
refer to Refs. \cite{SP1,SP2}. Furthermore, a calculation regarding
the multiplicative mass renormalization $Z_m$, which is directly
related to the flavor singlet scalar operator, can be found in
Ref. \cite{SP2}. The generalization of our results to an arbitrary
representation, as well as a detailed discussion regarding the
superficially divergent integrals, can also be found in these papers.
 \begin{eqnarray}
Z_V^{L,RI^{\,\prime}} = 1 &+& \frac{g_\circ^2}{16\pi^2}\,c_F
\Big{[}-20.617798655(6) + 4.745564682(3)\,c_{{\rm SW}} + 0.543168028(5)\,c_{{\rm SW}}^2 \Big{]} \nonumber \\
&+& \frac{g_\circ^4}{(16\pi^2)^2}\,c_F \Big{[} N_f\,\Big{(}25.610(3) -11.058(1)\,c_{{\rm SW}} + 33.937(3)\,c_{{\rm SW}}^2 \nonumber \\
&& \hskip 3cm -13.5286(6)\,c_{{\rm SW}}^3 -1.2914(6)\,c_{{\rm SW}}^4 \Big{)}  \nonumber \\
&&\qquad\qquad\quad + c_F\,\Big{(}-539.78(1) -223.57(2)\,c_{{\rm SW}} -104.116(5)\,c_{{\rm SW}}^2  \nonumber \\
&& \hskip 3.2cm -32.2623(8)\,c_{{\rm SW}}^3 +4.5575(3)\,c_{{\rm SW}}^4 \Big{)}  \nonumber \\
&&\qquad\qquad\quad + N_c\,\Big{(}-51.59(1) +18.543(5)\,c_{{\rm SW}} +20.960(6)\,c_{{\rm SW}}^2 \nonumber \\
&& \hskip 3.2cm +2.5121(5)\,c_{{\rm SW}}^3 +0.1765(1)\,c_{{\rm SW}}^4 \Big{)} \Big{]} 
\label{ZV2loopRI}
\end{eqnarray}
\begin{eqnarray}
Z_{AV}^{L,RI^{\,\prime}} = 1 &+& \frac{g_\circ^2}{16\pi^2}\,c_F \Big{[}
  -15.796283066(5) -0.247827627(3)\,c_{{\rm SW}} + 2.251366176(5)\,c_{{\rm SW}}^2 \Big{]} \nonumber \\
&+& \frac{g_\circ^4}{(16\pi^2)^2}\,c_F \Big{[} N_f\,\Big{(}18.497(3) -1.285(1)\,c_{{\rm SW}} + 19.071(3)\,c_{{\rm SW}}^2 \nonumber \\
&& \hskip 3cm +1.0333(6)\,c_{{\rm SW}}^3 -6.7549(6)\,c_{{\rm SW}}^4 \Big{)}  \nonumber \\
&&\hskip 2cm + c_F\,\Big{(}-184.01(1) -389.86(1)\,c_{{\rm SW}} -166.738(6)\,c_{{\rm SW}}^2  \nonumber \\
&& \hskip 3cm +7.894(1)\,c_{{\rm SW}}^3 + 4.3201(3)\,c_{{\rm SW}}^4 \Big{)}  \nonumber \\
&&\hskip 2cm + N_c\,\Big{(}-21.62(1) -33.652(5)\,c_{{\rm SW}} +26.636(6)\,c_{{\rm SW}}^2 \nonumber \\
&& \hskip 3cm +10.2186(5)\,c_{{\rm SW}}^3 +1.4893(1)\,c_{{\rm SW}}^4 \Big{)} \Big{]} 
\label{ZA2loopRI}
\end{eqnarray}

\end{document}